# Tunable phase-gradient-based optical tweezers

Xionggui Tang,*,†,‡ Fan Nan,§ and Zijie Yan*,§,†

†Department of Chemical and Biomolecular Engineering, Clarkson University, Potsdam, NY 13699, United States

‡Department of Physics, Key Laboratory of Low Dimensional Quantum Structures and Quantum Control of Ministry of Education, Synergetic Innovation Center for Quantum Effects and Applications, Hunan Normal University, Changsha, 410081, P.R. China

§Department of Applied Physical Sciences, University of North Carolina at Chapel Hill, Chapel Hill, NC 27599, United States

*E-mails: tangxg@hunnu.edu.cn (X.T.), zijieyan@unc.edu (Z.Y.)

**Abstract:** Conventional optical tweezers are generated by the intensity gradient of highly focused laser beams, but the requirement of strong intensity gradient limits the tunability of optical traps. Here we show a new type of optical tweezers with tunable potential wells by manipulating the phase gradient of light. Using a new method to calculate holograms, we can design desirable phase profiles and intensity distributions of optical patterns. Optical force arising from the phase gradient creates tunable potential wells for versatile optical manipulation, such as trapping nanoparticles in peanut-shaped optical spots, and positioning and shifting nanoparticles in optical gears as demonstrated in our experiments. The phase-gradient-based optical tweezers have several merits including flexible design and easy control, which open a new way for optical trapping and manipulation.

Optical tweezers have attracted increasing interest due to their wide applications in atomic physics, biology, chemical analysis and microfluidics.[1-6] Conventional optical tweezers rely on the intensity gradient forces exerted by strongly focused laser beams to trap various particles. Different manipulation techniques have also been demonstrated using optical tweezers, such as transporting, positioning, sorting and assembling micro- and nanoparticles.[7-9] However, the capability of traditional optical traps is limited by the requirement of strong intensity gradients. High intensity irradiation from laser easily leads to damage of samples, especially biological specimens.[10] In addition, their trapping potentials are generally static while tunable potential wells are highly desirable for optical trapping and manipulation. Optical tweezers with tunable potential wells can develop new functions and applications, such as dynamically manipulating particles, adjusting trapping stiffness, creating tunable trapping potentials, assembling novel functional microdevices, and effectively reducing photothermal effects.

Recently, several techniques for tuning the optical trapping potentials have been proposed based on fiber structures, waveguide structures, surface plasmons or holography.[11-19] Teeka et al. proposed a dynamic optical trap by using a dark soliton in an optical fiber loop to realize tunable widths and powers of optical tweezers,[11] and Mobini et al. tuned the optical force in a graded index multimode fiber based on the wavelength dependence of its numerical aperture.[12] Recently, Ping et al. demonstrated multifunctional manipulation by tunable optical lattices created by near-field modes beating along silicon waveguide.[13] In addition, several schemes have been presented to implement tunable optical forces based on surface plasmons. The surface plasmons can be controlled by adjusting illumination parameters such as optical polarization, wavelength or incident angle.[14-17] Tunable optical tweezers have also been achieved by creating structured optical patterns using space light modulators (SLMs).[18-19] However, all the aforementioned tunable optical traps are based on intensity gradient profiles. In 2008, Roichman et al. reported a new type of optical force arising from the phase gradient of light.[20] The optical force can create a potential barrier or well to trap silica microparticles in a flat-top line trap with a parabolic phase profile, and rotate them in a ring trap with orbital angular momentum. However, the shape-phase holography used to create the flat-top line trap suffers high laser power loss, which is highly unfavorable for optical manipulation of nanoparticles, and the ring traps can only continuously rotate particles with constant phase gradients.[21-22] So far tunable phase gradients have been mainly achieved in traditional optical line traps created by focusing a Gaussian beam with cylindrical lenses or the corresponding holograms,[23-25] yet in these optical lines the intensity and phase profiles are coupled that limits many their applications. It is a significant challenge to achieve versatile optical manipulation with optical forces coming from phase gradient.

In this letter, we report a new type of optical tweezers with tunable potential wells by controlling the phase gradient profiles. Optical traps with desired phase and intensity profiles at the plane of optical manipulation are generated by computer-generated holograms. Our experimental results show that the confinement of metal nanoparticles in optical traps can be easily controlled by adjusting the phase gradient. Furthermore, on-demand positioning, shifting, and transport of nanoparticles are demonstrated as examples of potential applications. The tunable phase-gradient-based optical tweezers have several features, including flexible design and control, which can significantly benefit optical nanomanipulation. This study thus provides a new way for developing novel functions of optical tweezers.

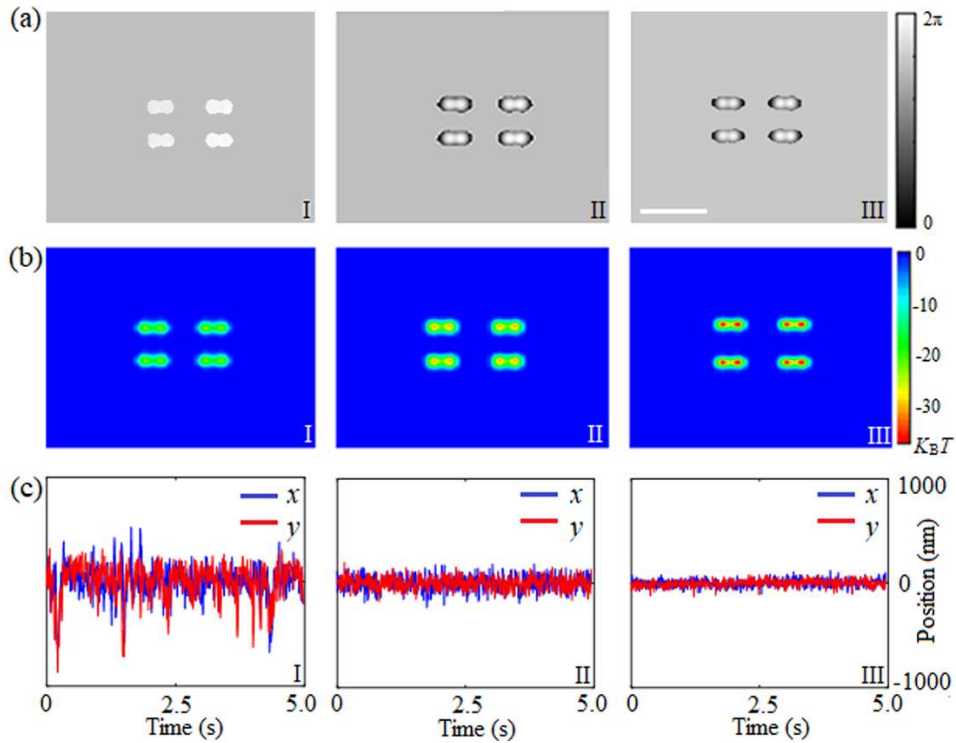

**Figure 1.** Peanut-shaped optical traps with tunable potential wells. (a) Phase profiles. (b) Potential energy distributions, where $k_B$ is Boltzmann's constant and $T = 300$ K. (c) The positions of the nanoparticles at different times. Note that panels I-III correspond to increasing phase gradients. The laser power used in experiment is 420 mW, and the scale bar is 5 μm.

In order to demonstrate phase-gradient-based optical tweezers, a holographic optical trapping system using a laser of 800 nm wavelength is employed to trap Au nanoparticles (see Methods in Supporting Information for experiment method). We use a one-step method to calculate phase-only holograms that can accurately generate optical traps with tunable phase gradient profiles at the Fourier plane of the holograms (see Methods in Supporting Information for hologram design method). We first demonstrate that different phase gradients can create variable potential wells by using peanut-shaped optical patterns with almost identical intensity distributions. The peanut-shaped spots are novel optical traps for realizing new functions, such as controllable alignment.[26] We investigate the stability of Au nanoparticles (200 nm diameter) in peanut-shaped spot optical traps with three different potential wells using different phase gradients. The holograms are first calculated, and their corresponding phase profiles and intensity profiles at the output plane are simulated. The simulated phase profiles are shown in Fig. 1a, whose phase gradients increase from panel I to III (for holograms, simulated intensity profiles, see Fig. S1 in supporting information). Their intensity distributions are almost identical, so the trapping potential wells caused be intensity distributions are nearly same. Fig. 1b shows the calculated trapping potential wells, which reveals that different potential wells are induced by different phase gradient profiles (see Methods in Supporting Information for the potential calculation method). Tunable potential wells have been created, and their trapping stiffnesses can be investigated by monitoring the fluctuation amplitudes of trapped nanoparticles. The time trajectories of $x$- and $y$-positions of Au nanoparticles trapped by different potential wells are plotted in Fig. 1c. It is worth noting that each peanut-shaped optical trap has two potential wells, so it has capability of trapping two or more nanoparticles. To avoid the influence of optical binding,[27] we here provide fluctuation analysis while only one nanoparticle is trapped in one of the potential wells. The standard deviations of fluctuation amplitudes are 157 nm, 72 nm and 36 nm in the $x$ direction, and are 176 nm, 61 nm and 33 nm in the $y$ direction, respectively. Obviously, larger phase gradient leads to smaller standard deviation, revealing that tuning the phase gradient can effectively alter the confinement of nanoparticles in optical traps. Moreover, the depths of measured potential wells, obtained from the position distribution probability of particles (see Methods and Fig. S2), are in good agreement with the calculated results in Fig. 1b, further demonstrating the validity of optical traps induced by phase gradient.

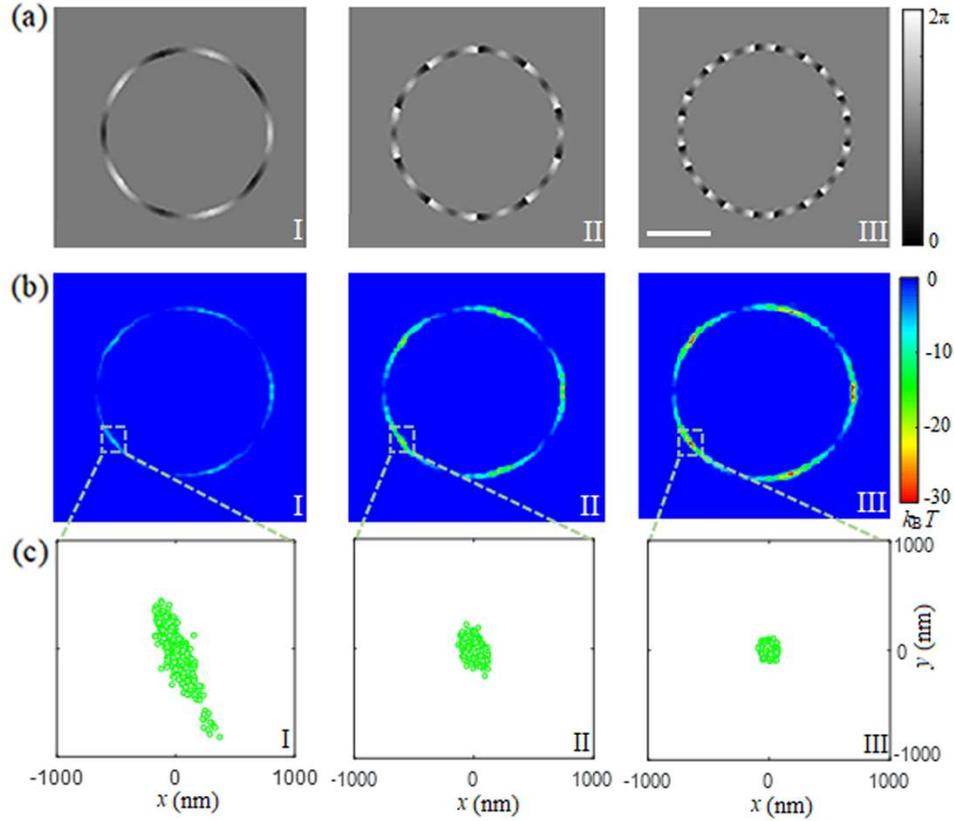

**Figure 2.** Ring traps with tunable potential wells. (a) Phase profiles. (b) Potential energy distributions. (c) The trajectories of a single particle trapped in a potential well. Panels I-III correspond to increasing phase gradients. The laser power is 191 mW, and the scale bar is 5 μm.

We further demonstrate tunable phase-gradient-based optical traps using optical rings and triangles with different phase gradients. Previously, different curved traps have been proposed to only drive particles moving in curved orbit[28]. However, positioning of nanoparticles by manipulating tunable potential wells has not been demonstrated yet. Here numerical simulations and related experiments are carried out with tunable ring traps. Their simulated phase profiles are given in Fig. 2a, where the phase gradient increases from left to right (for holograms and the simulated intensity profiles, see Fig. S3 in supporting information). Similarly, their intensity distributions are approximately identical, but the phase distributions are different, which easily causes different potential wells along a ring orbit as plotted in Fig. 2b. The trajectories of Au nanoparticles trapped by the potential wells are shown in Fig. 2c. Their standard deviations of fluctuations are 243 nm, 84 nm and 46 nm, respectively, which further demonstrates that the capability of particle confinement can be dynamically controlled by generating different potential wells based on phase gradient.

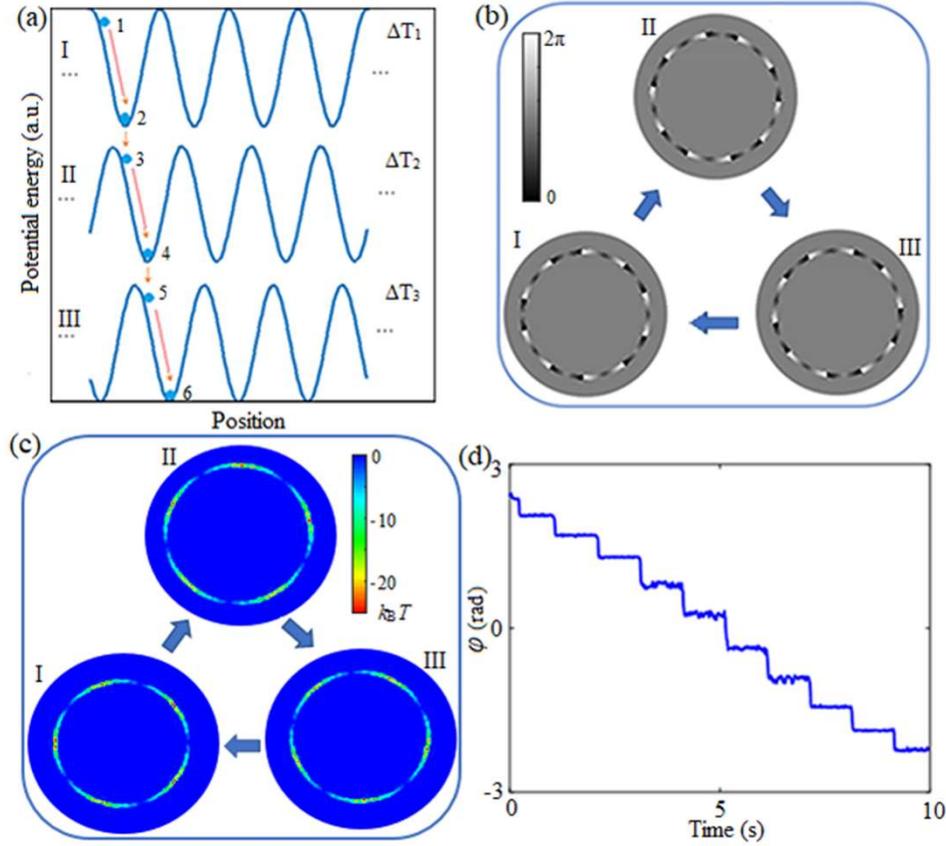

**Figure 3.** Controllable positioning and shifting of nanoparticles using an optical gear. (a) Principle of experiment. (b) The phase profiles at different times. (c) The potential wells at different times. (d) Angular trajectory of a nanoparticle in the ring orbit. The laser power is 315 mW.

The phase-gradient-based optical tweezers provide new opportunities for optical manipulation, for example, controllable positioning and shifting. Optical spot arrays are usually used to trap and shift nanoparticles or atoms,[29] but it is relatively slow to shift nanoparticles from one location to another one by moving spot trap arrays[30]. Here we demonstrate that the locations of potential wells generated by phase gradient can be intentionally shifted along a circle by dynamically varying its hologram, i.e., an optical gear. Fig. 3a shows the physical principle for controllable positioning and shifting of nanoparticles using an optical gear made with ring traps. Three different potential profiles (marked as I, II and Ш) are periodically employed for manipulating nanoparticles with time periods of $\Delta T_1$, $\Delta T_2$, and $\Delta T_3$, respectively. At the beginning, a nanoparticle is assumed to be located at position 1 with high potential energy in the first time period $\Delta T_1$, and it rapidly moves to position 2 with the lowest potential energy and keeps stable. Once the potential profile is switched from I to II, potential energy of the nanoparticle is instantly changed from low (position 2 on profile I) to high (position 3 on profile II), so it rapidly moves to position 4. As a result, the nanoparticle can be dynamically controlled to be positioned and shifted from one to another location. Its step distance, velocity, path, and time interval can be arbitrarily predesigned through holograms switched by LabView software. Fig. 3b presents their simulated phase profiles in different time interval (for holograms, simulated intensity profiles, see Fig. S4 in supporting information). Accordingly, the locations of potential wells can be easily changed by dynamically varying holograms, as shown in Fig. 3c. The positions of a trapped nanoparticle in the ring orbit are given in Fig. 3d. The nanoparticle can be quickly shifted from one to another location, and it can be well positioned in a desired place. Multiple nanoparticles can also be manipulated simultaneously. Specially, controllable positioning and shifting in a triangle orbit is feasible, indicating that the phase-gradient-based optical tweezers have potential application in optical manipulation such as controllable conveying and arrangement of nanoparticles. Additionally, the effect of intensity distribution on positioning and shifting of particles are discussed (see Methods in Supporting Information for effect of intensity distribution).

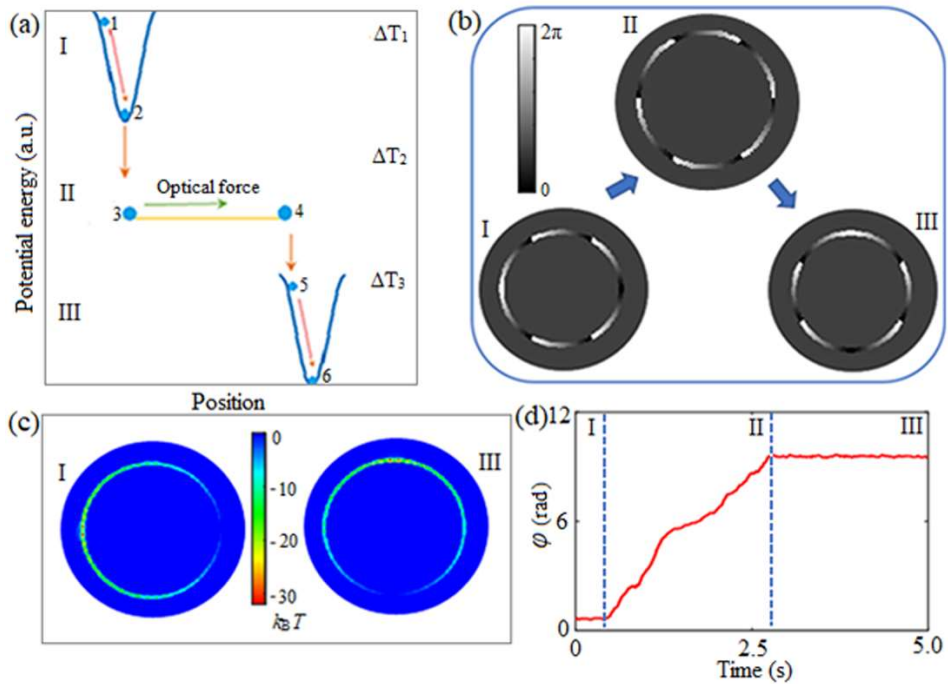

**Figure 4.** Controllable transport of nanoparticles. (a) Principle of experiment. (b) Phase profiles at different times. (c) Potential wells for the phase-gradient-based traps I and III. (d) Angular trajectory of a nanoparticle in the ring orbit. The laser power is 135 mW.

We further show that the tunable potential wells can be used for achieving controllable transport of nanoparticles by combining with conventional ring traps.[22] The experimental principle is briefly described in Fig. 4a, in which two potential wells (marked as I and III) are utilized for trapping, and a constant phase gradient profile (II) provides a constant optical force to drive a nanoparticle. First, a nanoparticle is assumed to be located at position 1, and then rapidly moves to position 2, which keeps stable at the bottom of potential well. When the hologram is switched from I to II, a conventional ring trap is adopted for driving the nanoparticle from position 3 to position 4 along a designed orbit. The hologram is switched again to create a potential well III when the nanoparticle arrives at position 4, so the nanoparticle will move from place 5 to place 6 and be trapped there. In this process, the start and end positions of nanoparticle, and its path and velocity along orbit can be purposely controlled by using the desired holograms. Fig. 4b presents the simulated phase profiles in different time periods (for holograms, simulated intensity profiles, see Fig. S5 in supporting information). Accordingly, the desired locations of potential wells can be easily manipulated, as shown in Fig. 4c. The controllable transport of a trapped Au nanoparticle in the ring orbit are plotted in Fig. 4d, and the nanoparticle can be controllably transported from the start position to the destination, where it keeps stable. Such optical manipulation can be used as a novel tool for microscopic cargo delivery.

In conclusion, we have proposed and experimentally demonstrated tunable optical tweezers by controlling the transverse phase gradient of light, which belong to 2D traps working against the coverslip. The phase-gradient-based optical tweezers provide a new way for generating various potential energy profiles, which has the potential for developing new functions and applications in optical trapping and manipulation. For instance, it can be used for tuning the confinement of nanoparticles, and positioning, shifting and transporting nanoparticles from one place to another. The ability to create a trapping potential by phase gradient largely expands the capacity of optical tweezers that so far mainly rely on the intensity gradient force. Generally, optical force arising from intensity gradient is used for trapping particles, while optical force exerted by phase gradient is adopted for driving particles in specific orbits, such as optical vortex and ring traps. In the future, however, different optical forces arising from the intensity and phase gradients can be individually generated or combined by smartly designing the holograms, which will greatly broaden the applications of optical tweezers for exploring novel function in optical manipulation.

■ ACKNOWGMENTS

This work was supported by the W. M. Keck Foundation Research Program. X.T. also thanks the Foundation of China Scholarship Council for support.

■ ASSOCIATED CONTENT

**Supporting Information**

The Supporting Information is available free of charge on the ACS Publications website at DOI:

Experimental and potential well calculation methods, and additional results (PDF)

■ REFERENCES

(1) Grier, D. G. A revolution in optical manipulation. *Nature* **2003**, 424, 810-816.
(2) Maragò, O. M.; Jones, P. H.; Gucciardi, P. G.; Volpe, G.; Ferrari, A. C. Optical trapping and manipulation of nanostructures. *Nat. Nanotech.* **2013,** 8, 807-819.
(3) Bowman, R. W.; Padgett, M. J. Optical trapping and binding. *Rep. Prog. Phys*. **2013,** 76, 026401(1-28).
(4) Paiè, P.; Zandrini, T.; Vázquez, R. M.; Osellame, R.; Bragheri, F. Particle manipulation by optical forces in microfluidic devices. *Micromachines* **2018**, 9, 200(1-28).
(5) Gao, D.; Ding, W.; Nieto-Vesperinas, M.; Ding, X.; Rahman, M.; Zhang, T.; Lim, C. T.; Qiu, C. W. Optical manipulation from the microscale to the nanoscale: fundamentals, advances and prospects. *Light Sci. Appl.* **2017**, 6, e17039(1-15).
(6) Barredo, D.; Lienhard, V.; Léséleuc, S. d.; Lahaye, T.; Browaeys, A. Synthetic three-dimensional atomic structures assembled atom by atom. *Nature* **2018**, 561, 79-82.

(7) Rodrigo, J. A.; Angulo, M.; Alieva, T. Programmable optical transport of particles in knot circuits and networks. *Opt. Lett.* **2018**, 43, 4244-4247.

(8) Nan, F.; Han, F.; Scherer, N. F.; Yan, Z. Dissipative self-assembly of anisotropic nanoparticle chains with combined electrodynamic and electrostatic interactions. *Adv. Mater.* **2018**, 30, 1803238(1-9).

(9) Tkachenko, G.; Brasselet, E. Optofluidic sorting of material chirality by chiral light. *Nat. Commun.* **2014**, 5, 3577(1-7).

(10) Huang, W. H.; Li, S. F.; Xu, H. T.; Xiang, Z. X.; Long, Y. B.; Deng, H. D. Tunable optical forces enhanced by plasmonic mode hybridization in optical trapping of gold nanorods with plasmonic nanocavity. *Opt. Express* **2018**, 26, 6202-6213.

(11) Teeka, C.; Jalil, M. A.; Yupapin, P. P.; Ali, J. Novel tunable dynamic tweezers using dark-bright soliton collision control in an optical add/drop filter. *IEEE T. Nanobiosci*. **2010**, 9, 258-262.

(12) Mobini, E.; Mafi, A. Design of a wavelength-tunable optical tweezer using a graded-index multimode optical fiber. *J. Lightwave Technol*. **2017**, 35, 3854-3861.

(13) Pin, C.; Jager, J. B.; Tardif, M.; Picard, E.; Hadji, E.; De Fornela, F.; Cluzel, B. Optical tweezing using tunable optical lattices along a few-mode silicon waveguide. *Lab Chip* **2018**, 18, 1750-1757.

(14) Righini, M.; Volpe, G.; Girard, C.; Petrov, D.; Quidant, R. Surface plasmon optical tweezers: tunable optical manipulation in the femtonewton range. *Phys. Rev. Lett*. **2008**, 100, 86804(1-4).

(15) Huang, W. H.; Li, S. F.; Xu, H. T.; Xiang, Z. X.; Long, Y. B.; Deng, H. D. Tunable optical forces enhanced by plasmonic mode hybridization in optical trapping of gold nanorods with plasmonic nanocavity. *Opt. Express* **2018**, 26, 6202-6213.

(16) Lu, Y.; Du, G.; Chen, F.; Yang, Q.; Bian, H.; Yong, J.; Hou, X. Tunable potential well for plasmonic trapping of metallic particles by bowtie nano-apertures. *Sci. Rep*. **2016**, 6, 32675(1-8).

(17) Huft, P. R.; Kolbow, J. D.; Thweatt, J. T.; Lindquist, N. C. Holographic plasmonic nanotweezers for dynamic trapping and manipulation. *Nano Lett*. **2017**, 17, 7920-7925.

(18) Bezryadina, A. S.; Preece, D. C.; Chen, J. C.; Chen, Z. Optical disassembly of cellular clusters by tunable 'tug-of-war' tweezers. *Light Sci. Appl*. **2016**, 5, e16158(1-7).

(19) Taylor, M. A. Optimizing phase to enhance optical trap stiffness. *Sci. Rep*. **2017**, 7, 555(1-10).

(20) Roichman, Y.; Sun, B.; Roichman, Y.; Amato-Grill, J.; Grier, D. G. Optical forces arising from phase gradients. *Phys. Rev. Lett*. **2008**, 100, 013602(1-4).

(21) Roichman, Y.; Grier, D. G. Projecting extended optical traps with shape-phase holography. *Opt. Lett.* **2006**, 31, 1675-1677.

(22) Figliozzi, P.; Sule, N.; Yan, Z.; Bao, Y.; Burov, S.; Gray, S. K.; Rice, S. A.; Vaikuntanathan, S.; Scherer, N. F. Driven optical matter: dynamics of electrodynamically coupled nanoparticles in an optical ring vortex. *Phys. Rev. E* **2017**, 95, 022604(1-14).

(23) Nan, F.; Yan, Z. Creating multifunctional optofluidic potential wells for nanoparticle manipulation. *Nano Lett.* **2018**, 18, 7400-7406.

(24) Yan, Z.; Sajjan, M.; Scherer, N. F. Fabrication of a material assembly of silver nanoparticles using the phase gradients of optical tweezers. *Phys. Rev. Lett*. **2015**, 114, 143901(1-5).

(25) Nan, F.; Yan, Z. Sorting metal nanoparticles with dynamic and tunable optical driven forces. *Nano Lett.* **2018**, 18, 4500-4505.

(26) David, G.; Esat, K.; Thanopulos, I.; Signorell, R. Digital holography of optically-trapped aerosol particles. *Commun. Chem*. **2018**, 1, 46(1-9).

(27) McCormack, P.; Han, F.; Yan, Z. Self-organization of metal nanoparticles in light: electrodynamics-molecular dynamics simulations and optical binding experiments. *J. Phys. Chem. Lett*. **2018**, 9, 545-549.

(28) Rodrigo, J. A.; Aueva, T. Freestyle 3D laser traps: tools for studying light-driven particle dynamics and beyond. *Optica* **2015**, 2, 812-815

(29) Endres, M.; Bernien, H.; Keesling, A.; Levine, H.; Anschuetz, E. R.; Krajenbrink, A.; Senko, C.; Vuletic, V.; Greiner, M.; Lukin, M. D. Atom-by-atom assembly of defect-free one-dimensional cold atom arrays. *Science* **2016**, 354, 1024-1027.

(30) Faucheux, L. P.; Bourdieu, L. S.; Kaplan, P. D.; Libchaber, A. J. Optical thermal ratchet, Physical Review Letters **1995**, 74,1504-1507.

# Supporting Information

# Tunable phase-gradient-based optical tweezers

Xionggui Tang,* Fan Nan, and Zijie Yan*

**METHODS**

**1. Hologram design method**

In the past two decades, various methods have been proposed, which includes the direct search algorithm[1], Gerchburg-Saxton algorithm[2], integral method[3]. Among them, the integral method can create high-accuracy intensity and phase profiles, but it is limited to the generation of smooth optical patterns made of special curves, called as superforma curves. Herein, we proposed a hologram calculation method[4], by introducing a random phase factor into the discrete inverse Fourier transform formula, which is a new approach for generating phase-only holograms that can produce high-quality intensity and phase profiles at the output plane. The approach is a direct computation method for CGH, which demonstrates that the calculation is very simple and computation time is low. Its formula is expressed by,

$$H(m,n) = c \sum_{\substack{k=-K/2 \\ s=-S/2}}^{\substack{k=K/2 \\ s=S/2}} U(k,s) \exp[j\frac{2\pi}{\lambda f}(km\Delta x_o \Delta x_i + sn\Delta y_o \Delta y_i) + j\phi], \qquad (S1)$$

where $c$ is a constant; $\Delta x_o$ and $\Delta y_o$ are the single pixel size in the $x$ and $y$ direction at the output plane; $\Delta x_i$ and $\Delta y_i$ are the single pixel size in the $x$ and $y$ direction at the input plane; $j$ is a complex unit, and $\lambda$ is operation wavelength and $f$ is focus length of objective lens; $k \in [-K/2, K/2]$, $s \in [-S/2, S/2]$, $m \in [-M/2, M/2]$, $n \in [-N/2, N/2]$, in which $K$ and $S$ stand for the maximum number of sampling pixels in the $x$ and $y$ axis at the output plane, respectively, and $M$, $N$ denote the maximum number of sampling pixels in the $x$ and $y$ axis at the input plane, respectively; $U(k,s)$ stands for $U(k\Delta x_o, s\Delta y_o)$, and $H(m,n)$ denotes $H(m\Delta x_i, n\Delta y_i)$; $U(k,s)$ is target optical field at back focus plane of objective lens; $H(m,n)$ is optical field reflected by SLM at front focus plane of objective lens; $\phi$ is a random phase factor, in which $\phi = d_m \cdot rand$, $d_m$ is the modulation depth with a maximum of 2π, and *rand* is a random function that varies in the range from 0 to 1. The introduction of random phase is to reduce the crosstalk among plane waves at the input plane, which can effectively improve quality of holograms. Consequently, the phase of hologram is written as,

$$P(m,n) = \arg[H(m,n)], \tag{S2}$$

where symbol arg stands for phase angle of $H(m,n)$. In this case, we can easily obtain holograms once the target optical field is given. In our design, each target pattern has 1024×1272 pixels, and its pixel size is 0.139 μm × 0.139 μm at the output plane. The phase-only SLM has 1024×1272 pixels with its pixel size of 12.5 μm × 12.5 μm. The focus length of the lens is 3 mm, and laser wavelength is 800 nm.

Additionally, we here provide the simulated results of rectangle-shaped optical patterns with linear phase gradient profiles, by using our method and the integral method, as shown in Fig. S6 (a-b), respectively. It reveals that the hologram designed by our method has high accuracy for generating intensity and phase profiles.

## 2. Experiment method

A phase-only SLM (Hamamatsu X13138), which has 1024×1272 pixels with its pixel size of 12.5 μm ×12.5 μm, is adopted to modulate the phase profiles of the designed hologram, and a CW tunable Ti:Sapphire laser (Spectra-Physics 3900s) operating at wavelength of 800 nm and producing a TEM00 Gaussian mode is employed, in which its optical polarization is along *x* direction. An inverted microscope (Olympus IX73) is used for generating optical traps, in which 60X objective (NA = 1.2, Olympus UPLSAPO 60XW) is employed. A beam profiler (Edmund Optics) is utilized for capturing the optical intensity of the reconstructed image. The laser powers used in Figs. 1- 4 are about 420 mW, 191 mW, 315 mW and 135 mW, respectively. The Au nanoparticles purchased from nanoComposix Inc. are nearly spherical, whose diameter is about 200 nm. The trapped Au nanoparticles are visualized by darkfield microscopy (Olympus U-DCW condenser), and recorded at frame rate of 150 fps, by using a CMOS camera (Point-Grey Grasshopper 3).

## 3. Potential well calculation

*From the designed phase profiles.* Generally, optical force resulting from a field with both intensity and phase gradients can be obtained by using the following formula,[5]

$$F = \frac{1}{4}\varepsilon_0\varepsilon\alpha'\nabla I + \frac{1}{2}\varepsilon_0\varepsilon\alpha''I\nabla\varphi, \quad (S3)$$

where $\varepsilon_0$ and $\varepsilon$ are dielectric constant and relative dielectric constant, $\alpha'$ and $\alpha''$ are the real and imaginary part of particle's polarizability, $I$ is the optical intensity, and $\varphi$ is the phase profile of light.

Next, potential energy surfaces are obtained by integrating the force along defined paths in the following formula,[6]

$$V(x,y) = -\int_{p_0}^{p} \vec{F} \cdot d\vec{l}, \quad (S4)$$

where $l$ denotes a given path from a reference point $p(x_0, y_0)$ to point $p(x, y)$ in x-y plane. The intensity and phase profiles are adopted from simulated results, which are the designed intensity and phase distributions. In this case, the light intensity is obtained by,

$$I = P/S \quad (S5)$$

where $P$ is the laser power and $S$ is the estimated area of optical trap. Consequently, the light intensity used for calculating potential wells in Fig. 1-4 are about 11 mW/μm², 11 mW/μm², 18 mW/μm² and 7.7 mW/μm², respectively. In addition, Au nanoparticles are illuminated by optical trappings in water. The refractive index of Au and water are 0.15+i4.91 and 1.33, respectively.

*From the experimental results.* The effective trapping potential of an optical trap can be calculated by the potential of mean force from the measured position (x and y) distribution of a trapped particle. The potential of mean force, *pmf(α)*, is given by

$$pmf(\alpha) = -k_B T \ln P(\alpha), \quad (S6)$$

where $k_B$ is Boltzmann's constant, $T$ is absolute temperature, and $P(\alpha)$ is the probability density of the parameter $\alpha$ (x, y).

## 4. Effect of intensity distribution

In our simulated results, the intensity distributions are almost identical. Consequently, it is reasonable that effect of intensity distribution and its related thermal consequence on particle trapping and shifting can be ignored. Experimentally, however, intensity distributions at trapping plane could be slightly deteriorated, i.e. the fluctuation of optical intensity will usually appear, which would lead to the different thermal distribution. These factors could have potential impact on particle trapping and shifting. Here, we take particle positioning and shifting as an example, and present a contrast experiment to investigate the effect of intensity distribution. In Fig. 3 and Fig. S4, an optical gear with phase gradient profiles is employed for controllable positioning and shifting of nanoparticles. Therefore, we design three corresponding holograms which generate the same intensity distribution, but zero phase gradients. The holograms, intensity distributions and phase profiles are shown in Fig. S7(a-c). Obviously, they are highly similar among the panels I-III in Fig. S7, because their phase gradients are all equal to zero. Followingly, the experiment for positioning and shifting is carried out by using holograms Fig S7(a), in which optical power is equal to one in Fig. 3. The position trajectory of a nanoparticle in the ring orbit is given in Fig. S8. It demonstrates that the particle motion in ring orbit belongs to random motion, which are caused by Brownian noise. Apparently, positioning and shifting of particle can't be realized in such a ring orbit where intensity distribution is almost identical and phase gradient is zero. Consequently, it is reasonable that the effect of intensity distribution and its related thermal consequence on particle trapping and shifting are ignored.

**FIGURES**

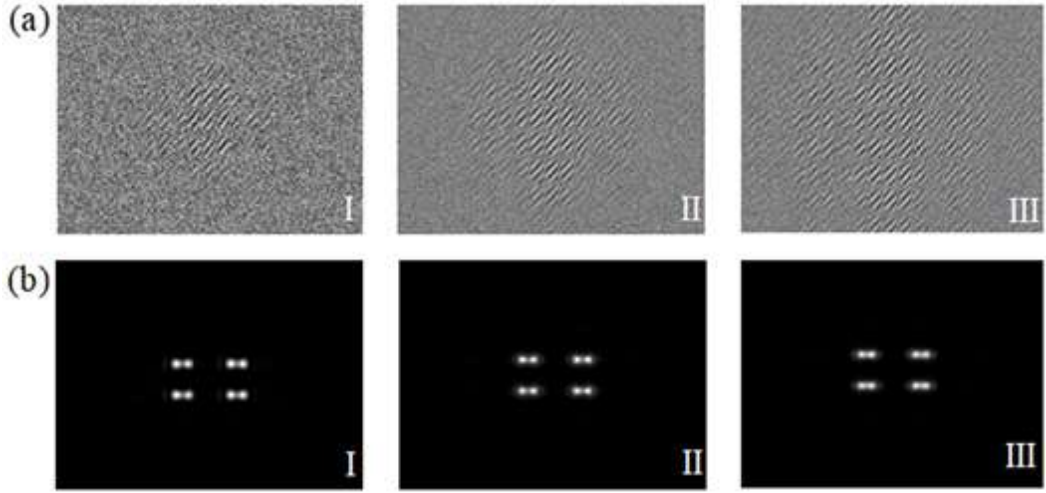

**Figure S1.** Peanut-shaped optical traps with tunable potential wells. (a) Holograms. (b) Simulated intensity profiles. Panels I-III correspond to increasing phase gradients. It is worth noting that their intensity distributions in Panels I-III are nearly identical.

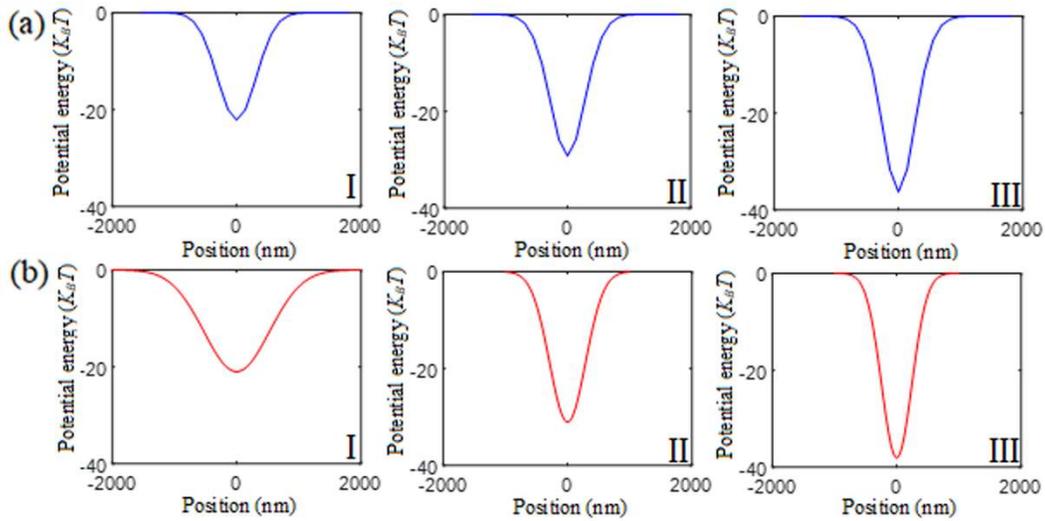

**Figure S2.** The potential energy curves. (a) The results from Fig. 1b. (b) The results obtained by using position probability of particles in Fig. 1c. Note that panels I- III correspond to increasing phase gradients, and potential energy curves along the $y$-axial direction of optical trap are provided, and the smallest potential energy is set at the origin by a coordinate system.

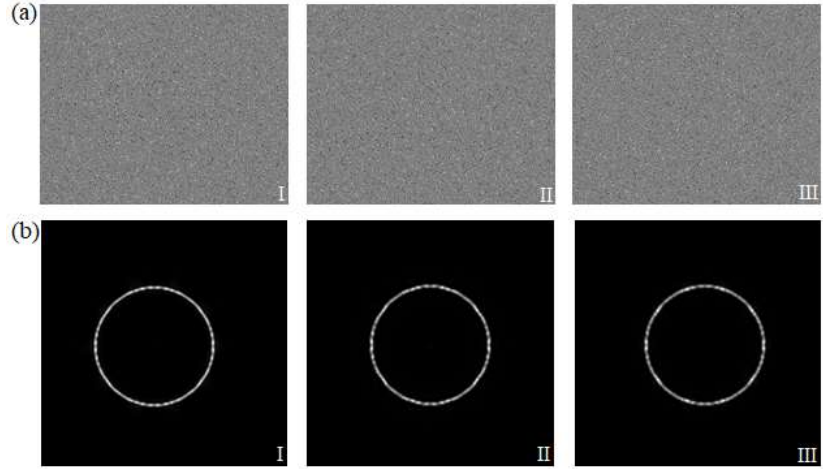

**Figure S3.** The ring optical traps with tunable potential wells. (a) Holograms. (b) Simulated intensity profiles. Note that panels I-III correspond to increasing phase gradients, but their intensity distributions in Panels I-III are almost identical.

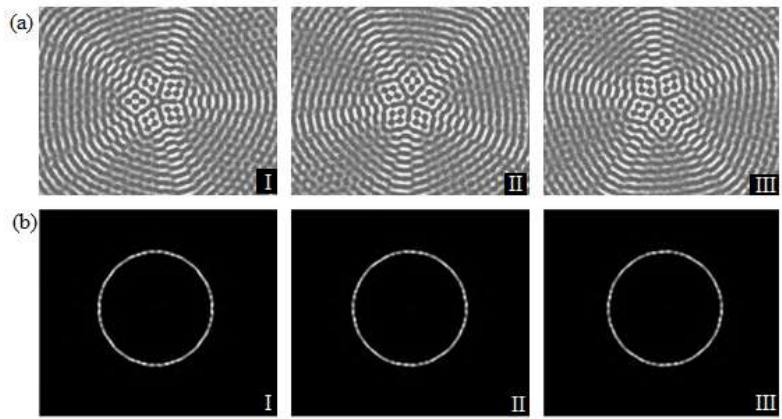

**Figure S4.** Controllable positioning and shifting of nanoparticles using an optical gear. (a) Holograms. (b) Simulated intensity profiles. Note that panels I-III correspond to different phase profile, but their intensity distributions in Panels I-III are almost identical.

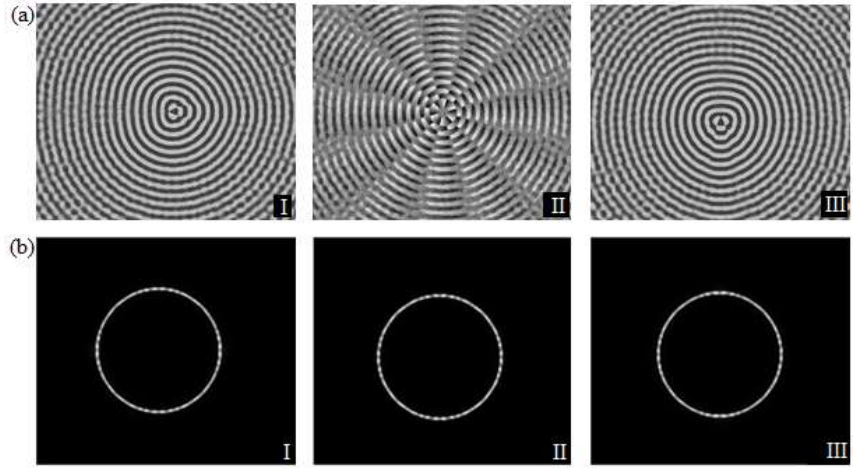

**Figure S5**. Controllable transport of nanoparticles. (a) Holograms. (b) Simulated intensity profiles. Note that panels I-III correspond to different phase profile, but their intensity distributions in Panels I-III are almost identical.

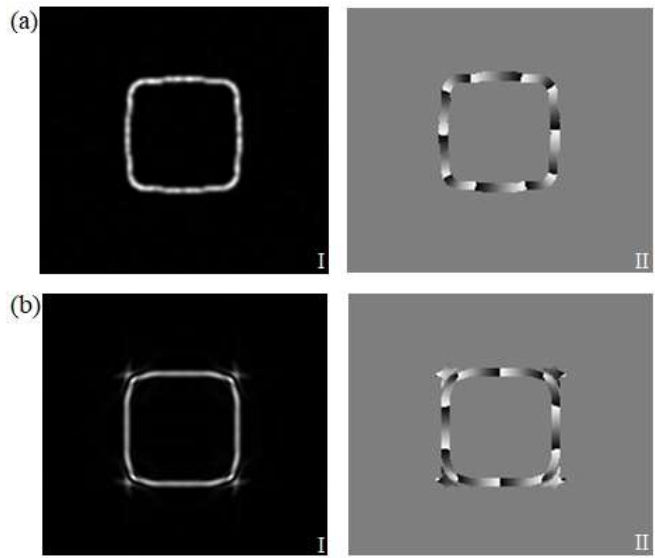

**Figure S6**. The simulated rectangle-shaped optical patterns by using (a) our method and (b) the integral method. Note that panels I- II correspond to intensity and phase profiles.

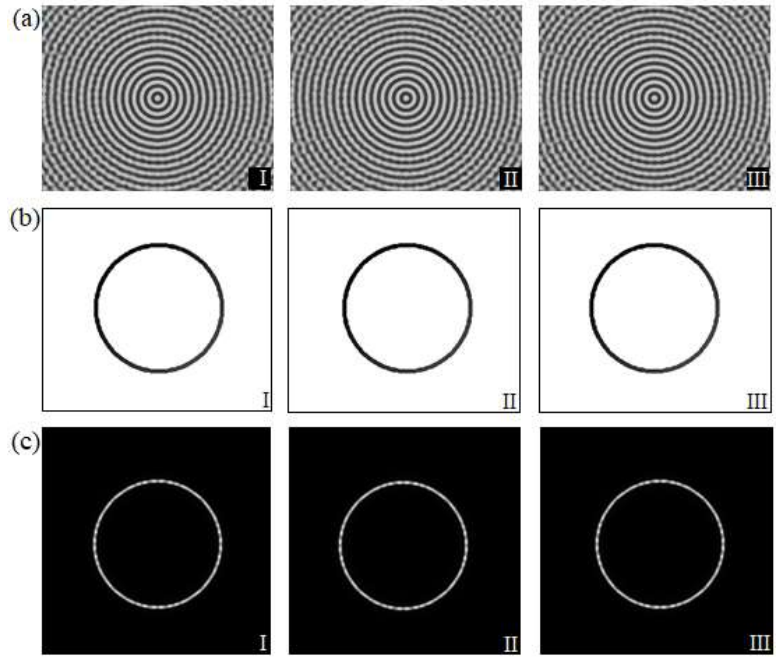

**Figure S7.** The effect of intensity distribution on positioning and shifting of nanoparticles. (a) Holograms. (b) Simulated phase profiles. (c) Simulated intensity profiles. Note that panels I-III correspond to ones in panels I-III of Fig. S4, but the phase gradients are all equal to zero.

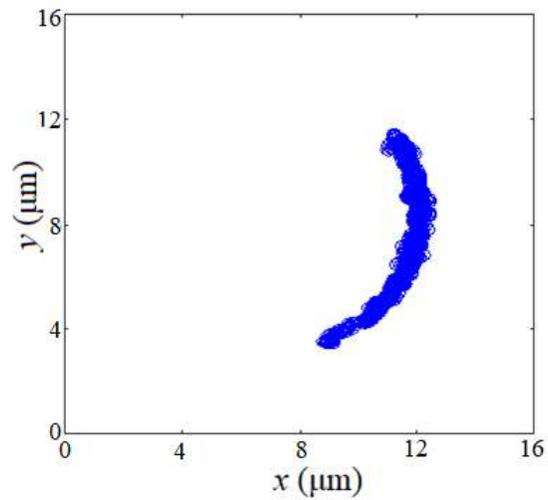

**Figure S8.** The position trajectories of a single particle in a ring orbit where intensity distribution is almost identical and phase gradient is zero.

**References**

(1) Polin, M.; Ladavac, K.; Lee, S.; Roichman, Y.; Grier, D. G. Optimized holographic optical traps. Optics Express 2005, 13, 5831-5845.

(2) Kim, H.; Kim, M.; Lee, W.; Ahn, J. Gerchberg-Saxton algorithm for fast and efficient atom rearrangement in optical tweezer traps. Optics Express 2019, 27, 2184-2196.

(3) Rodrigo, J. A.; Alieva, T. Freestyle 3D laser traps: tools for studying light-driven particle dynamics and beyond. Optica 2015, 2, 812-815.

(4) Tang, X.; Nan, F.; Han, F.; Yan, Z. Simultaneously shaping the intensity and phase of light for optical nanomanipulation, arXiv:1910.08244.

(5) Urban, A. S.; Carretero-Palacios, S.; Lutich, A. A.; Lohmüller, T.; Feldmann, J.; Jäckel, F. Optical trapping and manipulation of plasmonic nanoparticles: fundamentals, applications, and perspectives. *Nanoscale* **2014**, 6, 4458-4474.

(6) Yan, Z.; Gray, S. K.; Scherer, N. F. Potential energy surfaces and reaction pathways for light-mediated self-organization of metal nanoparticle clusters. *Nat. Commun*. **2014**, 5, 3751.